\documentclass[12pt,journal,draftcls,letterpaper,onecolumn]{IEEEtran}

\usepackage{amssymb,amsmath,amsfonts,bm,epsfig,graphicx,theorem,latexsym}
\usepackage{rotating,setspace,latexsym,epsf,color,subfigure}
\usepackage{cite}
\newcommand{\Ev}{\mathbf{E}}
\newcommand{\tv}{\mathbf{t}}

\newcommand{\rv}{\mathbf{r}}
\newcommand{\lv}{\mathbf{l}}

\newcommand{\pv}{\mathbf{p}}

\newtheorem{Theorem}{Theorem}
\newtheorem{Lemma}{Lemma}

\newenvironment{Proof}[1]{\medskip\par\noindent
{\bf Proof:\,}\,#1}{{\mbox{\,$\blacksquare$}\par}}

\makeatletter

\begin{document}
\IEEEoverridecommandlockouts
\pagestyle{empty}

\title{Optimal Packet Scheduling in an Energy Harvesting Communication System\thanks{This work was supported by NSF Grants CCF 04-47613, CCF 05-14846, CNS
07-16311, CCF 07-29127, CNS 09-64632 and presented in part at the $44$th Annual Conference on Information Sciences and Systems, Princeton, NJ, March 2010.}}

\author{Jing Yang \qquad Sennur Ulukus \\
\normalsize Department of Electrical and Computer Engineering \\
\normalsize University of Maryland, College Park, MD 20742 \\
\normalsize {\it yangjing@umd.edu} \qquad {\it ulukus@umd.edu} }

\maketitle \thispagestyle{empty}

\begin{abstract}
We consider the optimal packet scheduling problem in a single-user energy harvesting wireless communication system. In this system, both the data packets and the harvested energy are modeled to arrive at the source node randomly. Our goal is to adaptively change the transmission rate according to the traffic load and available energy, such that the time by which all packets are delivered is minimized. Under a deterministic system setting, we assume that the energy harvesting times and harvested energy amounts are known before the transmission starts. For the data traffic arrivals, we consider two different scenarios. In the first scenario, we assume that all bits have arrived and are ready at the transmitter before the transmission starts. In the second scenario, we consider the case where packets arrive during the transmissions, with known arrival times and sizes. We develop optimal off-line scheduling policies which minimize the time by which all packets are delivered to the destination, under causality constraints on both data and energy arrivals.
\end{abstract}

\newpage
\pagestyle{plain}
\setcounter{page}{1}
\pagenumbering{arabic}

\section{Introduction}
\begin{figure}[h]
\begin{center}
\scalebox{0.7} {\epsffile{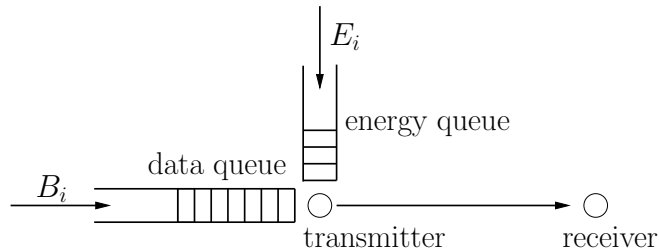}}
\end{center}
\caption{An energy harvesting communication system model.}
\label{fig:system1}
\vspace*{-0.15in}
\end{figure}

We consider wireless communication networks where nodes are able to harvest energy from nature. The nodes may harvest energy through solar cells, vibration absorption devices, water mills, thermoelectric generators, microbial fuel cells, etc. In this work, we do not focus on how energy is harvested, instead, we focus on developing transmission methods that take into account the {\it randomness} both in the {\it arrivals of the data packets} as well as in the {\it arrivals of the harvested energy}. As shown in Fig.~\ref{fig:system1}, the transmitter node has two queues. The data queue stores the data arrivals, while the energy queue stores the energy harvested from the environment. In general, the data arrivals and the harvested energy can be represented as two independent random processes. Then, the optimal scheduling policy becomes that of adaptively changing the transmission rate and power according to the instantaneous data and energy queue lengths.

While one ideally should study the case where both data packets and energy arrive randomly in time as two stochastic processes, and devise an {\it on-line} algorithm that updates the instantaneous transmission rate and power in {\it real-time} as functions of the current data and energy queue lengths, this, for now, is an intractable mathematical problem. Instead, in order to have progress in this difficult problem, we consider an idealized version of the problem, where we assume that we know exactly when and in what amounts the data packets and energy will arrive, and develop an optimal {\it off-line} algorithm. We leave the development of the corresponding {\it on-line} algorithm for future work.

\begin{figure}[t]
\begin{center}
\scalebox{0.7} {\epsffile{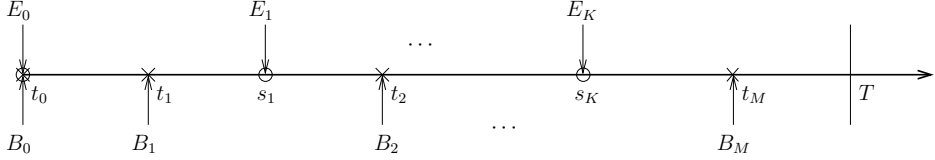}}
\end{center}
\caption{System model with random packet and energy arrivals. Data packets arrive at points denoted by $\times$ and energies arrive (are harvested) at points denoted by $\circ$.}
\label{fig:bit_arri}
\vspace*{-0.15in}
\end{figure}

Specifically, we consider a single node shown in Fig.~\ref{fig:bit_arri}. We assume that packets arrive at times marked with $\times$ and energy arrives (is harvested) at points in time marked with $\circ$. In Fig.~\ref{fig:bit_arri}, $B_i$ denotes the number of bits in the $i$th arriving data packet, and $E_i$ denotes the amount of energy in the $i$th energy arrival (energy harvesting). Our goal then is to develop methods of transmission to minimize the time, $T$, by which all of the data packets are delivered to the destination. The most challenging aspect of our optimization problem is the {\it causality} constraints introduced by the packet and energy arrival times, i.e., a packet may not be delivered before it has arrived and energy may not be used before it is harvested.

The trade-off relationship between delay and energy has been well investigated in traditional battery powered (unrechargeable) systems. References \cite{acm_2002, modiano_calculus, modiano_fading, ws_chen07, infocom_2002,it_2004} investigate energy minimization problems with various deadline constraints. Reference \cite{acm_2002} considers the problem of minimizing the energy in delivering all packets to the destination by a deadline. It develops a {\it lazy scheduling algorithm}, where the transmission times of all packets are equalized as much as possible, subject to the deadline and causality constraints, i.e., all packets must be delivered by the deadline and no packet may be transmitted before it has arrived. This algorithm also elongates the transmission time of each packet as much as possible, hence the name, {\it lazy} scheduling.
Under a similar system setting, \cite{modiano_calculus} proposes an interesting novel calculus approach to solve the energy minimization problem with individual deadlines for each packet. Reference \cite{modiano_fading} develops dynamic programming formulations and determines optimality conditions for a situation where channel gain varies stochastically over time. Reference \cite{ws_chen07} considers energy-efficient packet transmission
with individual packet delay constraints over a fading channel, and develops a recursive algorithm to find an optimal off-line schedule. This optimal off-line scheduler equalizes the energy-rate derivative function as much as possible subject to the deadline and
causality constraints. References \cite{infocom_2002} and \cite{it_2004} extend the single-user problem to multi-user scenarios. Under a setting similar to \cite{acm_2002}, we investigate the average delay minimization problem with a given amount of energy, and develop iterative algorithms and analytical solutions under various data arrival assumptions in \cite{jing_icc}. References \cite{berry01,shamai, sharma, yeh,javidi,yeh_mlqhr,musy} investigate delay optimal resource allocation problems under various different settings. References \cite{berry01,shamai, sharma}
consider average power constrained delay minimization problem for a single-user system, while \cite{yeh,javidi,yeh_mlqhr,musy} minimize the average delay through rate allocation in a multiple access channel.

In this paper, we consider a single-user communication channel with an energy harvesting transmitter. We assume that an initial amount of energy is available at $t=0$. As time progresses, certain amounts of energies will be harvested. While energy arrivals should be modeled as a random process, for the mathematical tractability of the problem, in this paper, we assume that the energy harvesting procedure can be precisely predicted, i.e., that, at the beginning, we know exactly when and how much energy will be harvested. For the data arrivals, we consider two different scenarios. In the first scenario, we assume that packets have already arrived and are ready to be transmitted at the transmitter before the transmission starts. In the second scenario, we assume that packets arrive during the transmissions. However, as in the case of energy arrivals, we assume that we know exactly when and in what amounts data will arrive. Subject to the energy and data arrival constraints, our purpose is to minimize the time by which all packets are delivered to the destination through controlling the transmission rate and power.

This is similar to the energy minimization problem in \cite{acm_2002}, where the objective is to minimize the energy consumption with a given {\it deadline} constraint. In this paper, minimizing the transmission completion time is akin to minimizing the deadline in \cite{acm_2002}. However, the problems are different, because, we do not know the exact amount of energy to be used in the transmissions, even though we know the times and amounts of harvested energy. This is because, intuitively, using more energy reduces the transmission time, however, using more energy entails waiting for energy arrivals, which increases the total transmission time. Therefore, minimizing the transmission completion time in the system requires a sophisticated utilization of the harvested energy. To that end, we develop an algorithm, which first obtains a good lower bound for the final total transmission duration at the beginning, and performs rate and power allocation based on this lower bound. The procedure works progressively until all of the transmission rates and powers are determined. We prove that the transmission policy obtained through this algorithm is globally optimum.

\section{Scenario I: Packets Ready Before Transmission Starts}
\label{sec:no_arri}
\begin{figure}[h]
\begin{center}
\scalebox{0.7} {\epsffile{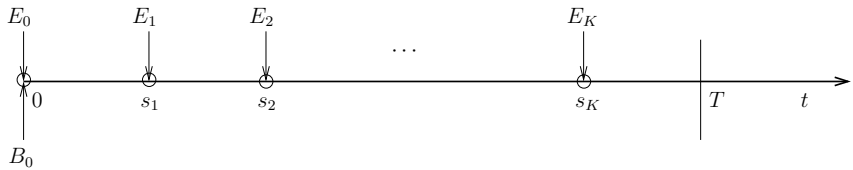}}
\end{center}
\caption{System model with all bits available at the beginning. Energies arrive at points denoted by $\circ$.}
\label{fig:system}
\vspace*{-0.15in}
\end{figure}

We assume that there are a total of $B_0$ bits available at the transmitter at time $t=0$. We also assume that there is $E_0$ amount of energy available at time $t=0$, and at times $s_1$, $s_2$, $\ldots$, $s_K$, we have energies harvested with amounts $E_1$, $E_2$, $\dots$, $E_K$, respectively. This system model is shown in Fig.~\ref{fig:system}. Our objective is to minimize the transmission completion time, $T$.

We assume that the transmitter can adaptively change its transmission power and rate according to the available energy and the remaining number of bits. We assume that the transmission rate and transmit power are related through a function, $g(p)$, i.e., $r=g(p)$. We assume that $g(p)$ satisfies the following properties: i) $g(0)=0$ and $g(p)\rightarrow \infty$ as $p\rightarrow \infty$, ii)
$g(p)$ increases monotonically in $p$, iii) $g(p)$ is strictly
concave in $p$,  iv) $g(p)$ is continuously differentiable, and v) $g(p)/p$ decreases monotonically in $p$. Properties i)-iii) guarantee that $g^{-1}(r)$ exists and is strictly convex. Property  v) implies that for a fixed amount of energy, the number of bits that can be transmitted increases as the transmission duration increases. It can be verified that these properties are satisfied in many systems with realistic encoding/decoding schemes, such as optimal random coding in single-user additive white Gaussian noise channel, where $g(p)=\frac{1}{2}\log(1+p)$.

\begin{figure}[t]
\begin{center}
\scalebox{0.7} {\epsffile{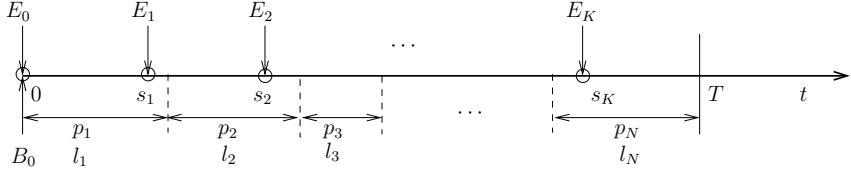}}
\end{center}
\caption{The sequence of transmission powers and durations.}
\label{fig:system2}
\vspace*{-0.15in}
\end{figure}

Assuming the transmitter changes its transmission power $N$ times before it finishes the transmission, let us denote the sequence of transmission powers as $p_1$, $p_2$, $\ldots$, $p_N$, and the corresponding transmission durations of each rate as $l_1$, $l_2$, $\ldots$, $l_N$, respectively; see Fig.~\ref{fig:system2}. Then, the energy consumed up to time $t$, denoted as $E(t)$, and the total number of bits departed up to time $t$, denoted as $B(t)$, can be related through the function $g$ as follows:
\begin{align}
E(t)&=\sum_{i=1}^{\bar{i}}p_il_i+p_{\bar{i}+1}\left(t-\sum_{i=1}^{\bar{i}}l_i\right)
\label{eqn:E}\\
B(t)&=\sum_{i=1}^{\bar{i}}g(p_i)l_i+g(p_{\bar{i}+1})\left(t-\sum_{i=1}^{\bar{i}}l_i\right)
\label{eqn:D}
\end{align}
where $\bar{i}=\max \{i: \sum_{j=1}^{i}l_j\leq t\}$.

Then, the transmission completion time minimization problem can be formulated as:
\begin{eqnarray}
\min_{\pv,\lv} & &T \nonumber \\
\mbox{s.t.} & & E(t)\leq\sum_{i:s_i< t} E_i, \quad  \quad 0\leq t\leq T \nonumber \\
& & B(T)=B_0
\label{opt_prob}
\end{eqnarray}
First, we determine the properties of the optimum solution in the following three lemmas.

\begin{Lemma}\label{lemma:incrs}
Under the optimal solution, the transmit powers increase monotonically, i.e., $p_1\leq p_2 \leq \cdots \leq p_N$.
\end{Lemma}
\begin{Proof}
Assume that the powers do not increase monotonically, i.e., that we can find two powers such that $p_{i}>p_{i+1}$. The total energy consumed over this duration is $p_il_i+p_{i+1}l_{i+1}$. Let
\begin{align}
    p_i'&=p'_{i+1}=\frac{p_il_i+p_{i+1}l_{i+1}}{l_i+l_{i+1}}\\
    r_i'&=r'_{i+1}=g\left(\frac{p_il_i+p_{i+1}l_{i+1}}{l_i+l_{i+1}}\right)
\end{align}
Then, we have $ p_i'\leq p_i$, $p'_{i+1}\geq p_{i+1}$. Since $p_i'l_i\leq p_il_i$, the energy constraint is still satisfied, and thus, the new energy allocation is feasible.  We use $ r_i', r'_{i+1}$ to replace $r_i, r_{i+1}$ in the transmission policy, and keep the rest of the rates the same. Then, the total number of bits transmitted over the duration $l_i+l_{i+1}$ becomes
\begin{align}
 r_i'l_i+r_{i+1}'l_{i+1}
&=g\left(\frac{p_il_i+p_{i+1}l_{i+1}}{l_i+l_{i+1}}\right)(l_i+l_{i+1})\nonumber\\
&\geq g\left(p_i\right)\frac{l_i}{l_i+l_{i+1}}(l_i+l_{i+1})+g\left(p_{i+1}\right)\frac{l_{i+1}}{l_i+l_{i+1}}(l_i+l_{i+1})\nonumber\\
    &=r_il_i+r_{i+1}l_{i+1}\label{eqn:incr}
\end{align}
where the inequality follows from the fact that $g(p)$ is concave in $p$.
Therefore, the new policy departs more bits by time $\sum_{j=1}^{i+1}l_j$. Keeping the remaining transmission rates the same, the new policy will finish the entire transmission over a shorter duration. Thus, the original policy could not be optimal. Therefore, the optimal policy must have monotonically increasing powers (and rates).
\end{Proof}

\begin{figure}[t]
\begin{center}
\scalebox{0.7} {\epsffile{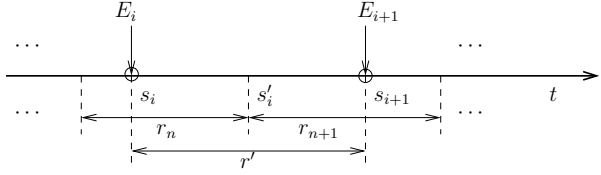}}
\end{center}
\caption{The rate must remain constant between energy harvests.}
\label{fig:const_rate}
\vspace*{-0.15in}
\end{figure}

\begin{Lemma}\label{lemma:const}
The transmission power/rate remains constant between energy harvests, i.e., the power/rate only potentially changes when new energy arrives.
\end{Lemma}
\begin{Proof}
Assume that the transmitter changes its transmission rate between two energy harvesting instances $s_i$, $s_{i+1}$. Denote the rates as $r_n$, $r_{n+1}$, and the instant when the rate changes as $s_i'$, as shown in Fig.~\ref{fig:const_rate}. Now, consider the duration $[s_i,s_{i+1})$. The total energy consumed during the duration is $p_n(s_i'-s_i)+p_{n+1}(s_{i+1}-s_i')$.  Let
   \begin{align}
   p'&=\frac{p_n(s_i'-s_i)+p_{n+1}(s_{i+1}-s_i')}{s_{i+1}-s_i}\\
   r'&=g\left(\frac{p_n(s_i'-s_i)+p_{n+1}(s_{i+1}-s_i')}{s_{i+1}-s_i}\right)
   \end{align}
Now let us use $r'$ as the new transmission rate over $[s_i,s_{i+1})$, and keep the rest of the rates the same. It is easy to check that the energy constraints are satisfied under this new policy, thus this new policy is feasible. On the other hand, the total number of bits departed over this duration under this new policy is
\begin{align}
  r'(s_{i+1}-s_i) 
  &=g\left(\frac{p_n(s_i'-s_i)+p_{n+1}(s_{i+1}-s_i')}{s_{i+1}-s_i}\right)(s_{i+1}-s_i)\nonumber\\
    &\geq \left(g(p_n)\frac{s_i'-s_i}{s_{i+1}-s_i}+g(p_{n+1})\frac{s_{i+1}-s_i'}{s_{i+1}-s_i}\right)(s_{i+1}-s_i)\nonumber\\
    &=r_n(s_i'-s_i)+r_{n+1}(s_{i+1}-s_i')
  \end{align}
where the inequality follows from the fact that $g(p)$ is concave in $p$. Therefore, the total number of bits departed under the new policy is larger than that under the original policy. If we keep all of the remaining rates the same, the transmission will be completed at an earlier time. This conflicts with the optimality of the original policy.
\end{Proof}

\begin{figure}[h]
\begin{center}
\scalebox{0.5} {\epsffile{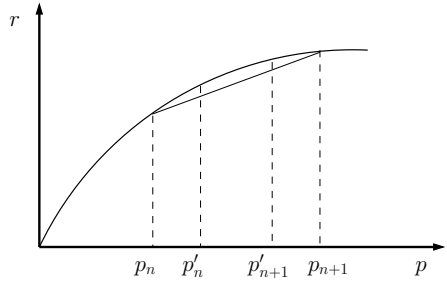}}
\end{center}
\caption{$g(p)$ is concave in $p$.}
\label{fig:concave}
\vspace*{-0.15in}
\end{figure}

\begin{Lemma}\label{lemma:energy}
Whenever the transmission rate changes, the energy consumed up to that instant equals the energy harvested up to that instant.
\end{Lemma}
\begin{Proof}
From Lemma~\ref{lemma:const}, we know that the transmission rate can change only at certain energy harvesting instances. Assume that the transmission rate changes at $s_i$, however, the energy consumed by $s_i$, which is denoted by $E(s_i)$, is less than $\sum_{j=0}^{i-1}E_j$. We denote the energy gap by $\Delta$. Let us denote the rates before and after $s_i$ by $r_{n}$, $r_{n+1}$. Now, we can always have two small amounts of perturbations $\delta_n$, $\delta_{n+1}$ on the corresponding transmit powers, such that
\begin{align}
    p'_n&=p_n+\delta_n\\
    p'_{n+1}&=p_{n+1}-\delta_{n+1}\\
    \delta_nl_n&=\delta_{n+1} l_{n+1}
\end{align}

We also make sure that $\delta_n$ and $\delta_{n+1}$ are small enough such that $\delta_nl_n<\Delta$, and $p'_n\leq p_{n+1}'$. If we keep the transmission rates over the rest of the duration the same, under the new transmission policy, the energy allocation will still be feasible. The total number of bits departed over the duration $(\sum_{i=1}^{n-1}l_i, \sum_{i=1}^{n+1}l_i)$ is
\begin{align}
    g(p'_{n})l_n+g(p'_{n+1})l_{n+1}\geq  g(p_{n})l_n+g(p_{n+1})l_{n+1}
\end{align}
where the inequality follows from the concavity of $g(p)$ in $p$, and the fact that $p_nl_n+p_{n+1}l_{n+1}=p'_nl_n+p'_{n+1}l_{n+1}$, $p_n\leq p_n'\leq p_{n+1}'\leq p_{n+1}$, as shown in Fig.~\ref{fig:concave}. This conflicts with the optimality of the original allocation.
\end{Proof}

We are now ready to characterize the optimum transmission policy. In order to simplify the expressions, we let $i_0=0$, and let $s_{m+1}=T$ if the transmission completion time $T$ lies between $s_{m}$ and $s_{m+1}$.

\begin{figure}[t]
\begin{center}
\scalebox{0.7} {\epsffile{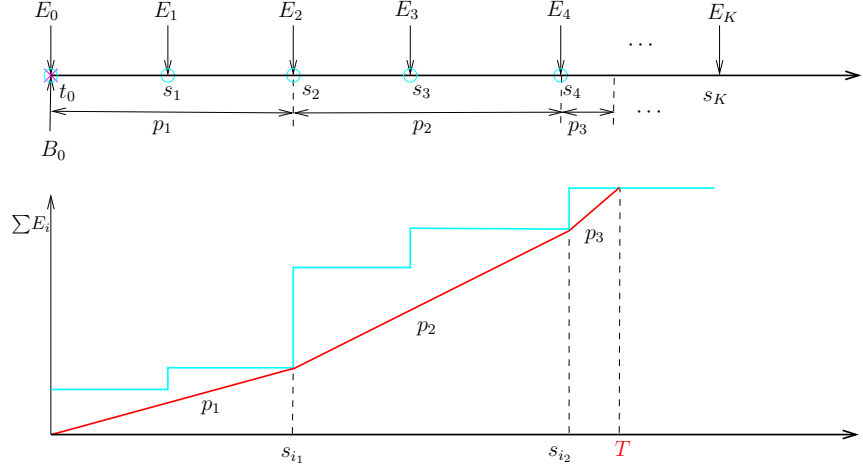}}
\end{center}
\caption{An interpretation of transmission policies satisfying Lemmas 1, 2 and 3.}
\label{fig:thm1}
\vspace*{-0.15in}
\end{figure}

Based on Lemmas~\ref{lemma:incrs}, \ref{lemma:const} and \ref{lemma:energy}, we can characterize the optimal policy in the following way. For given energy arrivals, we plot the total amount of harvested energy as a function of $t$, which is a staircase curve as shown in Fig.~\ref{fig:thm1}. The total energy consumed up to time $t$ can also be represented as a continuous curve, as shown in Fig.~\ref{fig:thm1}. In order to satisfy the feasibility constraints on the energy, energy consumption curve must lie below the energy harvesting curve at all times. Based on Lemma~\ref{lemma:const}, we know that the optimal energy consumption curve must be linear between any two consecutive energy harvesting instants, and the slope of the segment corresponds to the transmit power level during that segment. Lemma~\ref{lemma:energy} implies that whenever the slope changes, the energy consumption curve must touch the energy harvesting curve at that energy harvesting instant. Therefore, the first linear segment of the energy consumption curve must be one of the lines connecting the origin and any corner point on the energy harvesting curve before $t=T$. Because of the monotonicity property of the power given in Lemma~\ref{lemma:incrs}, among those lines, we should always pick the one with the minimal slope, as shown in Fig.~\ref{fig:thm1}. Otherwise, either the feasibility constraints on the energy will not be satisfied, or the monotonicity property given in Lemma 1 will be violated. For example, if we choose the line ending at the corner point at $s_3$, this will violate the feasibility constraint, as the energy consumption curve will surpass the energy arrival curve. On the other hand, if we choose the line ending at the corner point at $s_1$, then the monotonicity property in Lemma~\ref{lemma:incrs} will be violated, because in that case, the slope of the following segment would be smaller. These properties must hold similarly for $p_2$, $p_3$, $\ldots$, $p_N$. We also observe that, for given $T$, the optimal transmission policy is the tightest string below the energy harvesting curve connecting the origin and the total harvested energy by time $T$. This is similar to the structure in \cite{modiano_calculus}.

We state the structure of the optimal policy formally in the following theorem.
\begin{Theorem}\label{thm:structure}
For a given $B_0$, consider a transmission policy with power vector $\pv=[p_1, p_2,\ldots,p_N]$ and corresponding duration vector $\lv=[l_1,l_2,\ldots,\l_N]$. This policy is optimal if and only if it has the following structure:
\begin{align}
    \sum_{n=1}^N g(p_n)l_n&=B_0
    \end{align}
 and  for $n=1,2,\ldots, N$,
    \begin{align}
    i_n&= \arg \min_{\substack{i:
      s_i\leq T\\ s_i>s_{i_n-1}}}
    \left\{\frac{\sum^{i-1}_{j=i_{n-1}}E_j}{s_i-s_{i_{n-1}}}\right\}\\
    p_n&=\frac{\sum^{i_n-1}_{j=i_{n-1}}E_j}{s_{i_n}-s_{i_{n-1}}}\label{eqn:p_min}\\
    l_n&=s_{i_n}-s_{i_{n-1}}
\end{align}
where $i_n$ is the index of the energy arrival epoch when the power $p_n$ switches to $p_{n+1}$, i.e., at $t=s_{i_n}$, $p_n$ switches to $p_{n+1}$.
\end{Theorem}
The proof of this theorem is given in Appendix~\ref{apdix:thm1}.

Therefore, we conclude that if the overall transmission duration $T$ is known, then the optimal transmission policy is known via Theorem~\ref{thm:structure}. In particular, optimal transmission policy is the one that yields the tightest piecewise linear energy consumption curve that lies under the energy harvesting cure at all times and touches the energy harvesting curve at $t=T$. On the other hand, the overall transmission time $T$ is what we want to minimize, and we do not know its optimal value up front. Consequently, we do not know up front which energy harvests will be utilized. For example, if the number of bits is small, and $E_0$ is large, then, we can empty the data queue before the arrival of $E_1$, thus, the rest of the energy arrivals are not necessary. Therefore, as a first step, we first obtain a good lower bound on the optimal transmission duration.

\begin{figure}[t]
\begin{center}
\scalebox{0.7} {\epsffile{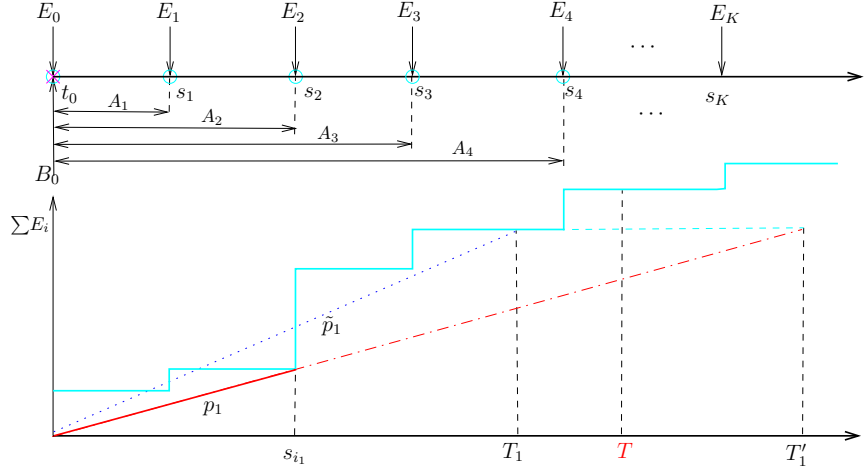}}
\end{center}
\caption{An illustration of the algorithm.}
\label{fig:proof2}
\vspace*{-0.15in}
\end{figure}

We first illustrate our algorithm through an example in Fig.~\ref{fig:proof2}. We first compute the minimal energy required to finish the transmission before $s_1$. We denote it as $A_1$, and it equals
\begin{align}
  A_1&=g^{-1}\left(\frac{B_0}{s_1}\right)s_1
\end{align}
Then, we compare it with $E_0$. If $A_1<E_0$, it implies that we can complete the transmission before the arrival of the first energy harvest, thus $E_1$ is not necessary for the transmission. We allocate $E_0$ evenly to $B_0$ bits, and the duration $A_1$ is the minimum transmission duration. On the other hand, if $A_1>E_0$, which is the case in the example, the final transmission completion time should be longer than $s_1$. Thus, we move on and compute $A_2$, $A_3$, $A_4$, and find that $A_2>\sum_{i=0}^1 E_i$, $A_3>\sum_{i=0}^2E_i$ and $A_4<\sum_{i=0}^3E_i$. This means that the total transmission completion time will be larger than $s_3$ and energies $E_0$, $\ldots$, $E_3$ will surely be utilized. Then, we allocate $\sum_{i=0}^3E_i$ evenly to $B_0$ bits and obtain a constant transmission power $\tilde{p}_1$, which is the dotted line in the figure. The corresponding transmission duration is $T_1$. Based on our allocation, we know that the final optimal transmission duration $T$ must be greater than $T_1$. This is because, this allocation assumes that all $E_0$, $\ldots$, $E_3$ are available at the beginning, i.e., at time $t=0$, which, in fact, are not. Therefore, the actual transmission time will only be larger. Thus, $T_1$ is a lower bound for $T$.

Next, we need to check the feasibility of $\tilde{p}_1$. Observing the figure, we find that $\tilde{p}_1$ is not feasible since it is above the staircase energy harvesting curve for some duration. Therefore, we connect all the corner points on the staircase curve before $t=T_1$ with the origin, and find the line with the minimum slope among those lines. This corresponds to the red solid line in the figure. Then, we update $\tilde{p}_1$ with the slope $p_1$, and the duration for $p_1$ is $l_1=s_{i_1}$. We repeat this procedure at $t=s_{i_1}$ and obtain $p_2$, and continue the procedure until all of the bits are finished.

We state our algorithm for the general scenario as follows:
First, we compute the amounts of energy required to finish the entire transmission before $s_1$, $s_2$, $\ldots$, $s_K$, respectively, at a constant rate. We denote these as $A_i$:
\begin{align}
  A_i&=g^{-1}\left(\frac{B_0}{s_i}\right)s_i, \quad i=1,2,\ldots,K
\end{align}
Then, we compare $A_i$ with $\sum_{j=0}^{i-1}E_j$, and find the smallest $i$ such that $A_i\leq \sum_{j=0}^{i-1}E_j$. We denote this $i$ as $\tilde{i}_1$. If no such $\tilde{i}_1$ exists, we let $\tilde{i}_1=K+1$.

Now, we assume that we can use $\sum_{j=0}^{\tilde{i}_1-1}E_j$ to transmit all $B_0$ bits at a constant rate. We allocate the energy evenly to these bits, and the overall transmission time $T_1$ is the solution of
\begin{align}
 g\left(\frac{\sum_{j=0}^{\tilde{i}_1-1}E_j}{T_1}\right)T_1&=B_0
\end{align}
and the corresponding constant transmit power is
\begin{align}
 p_1&=\frac{\sum_{j=0}^{\tilde{i}_1-1}E_j}{T_1}
\end{align}
Next, we compare $p_1$ with $\frac{\sum_{j=0}^{i-1}E_j}{s_i}$ for every $i<\tilde{i}_1$. If $p_1$ is smaller than every term, then, maintaining $p_1$ is feasible, and the optimal policy is to transmit at a constant transmission rate $g(p_1)$ with duration $T_1$, which gives the smallest possible transmission completion time,  $s_{i_1}=s_{\tilde{i}_1}$. Otherwise, maintaining $p_1$ is infeasible under the given energy arrival realization. Thus, we update
\begin{align}
  i_1&= \arg \min_{i<\tilde{i}_1} \left\{\frac{\sum_{j=0}^{i-1}E_j}{s_i}\right\}\\
  p_1&=\frac{\sum_{j=0}^{i_1-1}E_j}{s_{i_1}}
\end{align}
i.e., over the duration $[0, s_{i_1})$, we choose to transmit with power $p_1$ to make sure that the energy consumption is feasible. Then, at time $t=s_{i_1}$, the total number of bits departed is $g(p_1)s_{i_1}$, and the remaining number of bits is $B_0-g(p_1)s_{i_1}$. Subsequently, with initial number of bits $B_0-g(p_1)s_{i_1}$, we start from $s_{i_1}$, and get another lower bound on the overall transmission duration $T_2$, and repeat the procedure above. Through this procedure, we obtain $p_2,p_3,\ldots, p_N$, and the corresponding $i_2,i_3,\ldots, i_N$, until we finish transmitting all of the bits.

Based on our allocation algorithm, we know that $p_1$ is optimum up to time $T_1$, since it corresponds to the minimal slope line passing through the origin and any corner point before $t=T_1$. However, the algorithm also implies that the final transmission duration $T$ will be larger than $T_1$. The question then is, whether $p_1$ is still the minimum slope line up to time $T$. If we can prove that $p_1$ is lower than the slopes of the lines passing through the origin and any corner point in $[T_1, T]$, then, using Theorem~\ref{thm:structure}, we will claim that $p_1$ is the optimal transmission policy, not only between $[0, T_1]$, but also between $[0,T]$.

The fact that this will be the case can be illustrated through the example in Fig.~\ref{fig:proof2}. We note that, clearly, $T_1$ is a lower bound on the eventual $T$. If we keep transmitting at power $p_1$, if no additional energy arrives, the energy harvested up until $s_{\tilde{i}_1}$, i.e., $\sum_{i=0}^{\tilde{i}_1-1}E_i$, will be depleted by time $T_1'$. We will next prove that $T_1'$ is an upper bound on $T$. Because of the concavity of the function $g(p)$ in $p$, we can prove that under this policy, the number of bits departed up to time $T_1'$ is greater than $B_0$. Therefore, since potentially additional energy will arrive, $T_1'$ provides an upper bound. Thus, we know that the optimal $T$ lies between $T_1$ and $T_1'$. We next note that if we connect the origin with any corner point of the staircase curve between $T_1$ and $T_1'$, the slope of the resulting line will be larger than $p_1$, thus, $p_1$ will be the smallest slope not only up to time $T_1$, which is a lower bound, but also up to time $T_1'$, which is an upper bound. This proves that while we do not know the optimal $T$, if we run the algorithm with respect to the lower bound on $T$, i.e., $T_1$, it will still yield an optimal policy, in that the resulting policy will satisfy Theorem~\ref{thm:structure}.

We prove the optimality of the algorithm formally in the following theorem.

\begin{Theorem}\label{thm2}
The allocation procedure described above gives the optimal transmission policy.
\end{Theorem}
The proof of this theorem is given in Appendix~\ref{apdix:thm2}.

\section{Scenario II: Packets Arrive during Transmissions}
In this section, we consider the situation where packets arrive during transmissions. We assume that there is an $E_0$ amount of energy available at time $t=0$, and at times $s_1$, $s_2$, $\ldots$, $s_K$, energy is harvested in amounts $E_1$, $E_2$, $\dots$, $E_K$, respectively, as in the previous section. We also assume that at $t=0$, we have $B_0$ bits available, and at times $t_1$, $t_2$, $\ldots$, $t_M$, bits arrive in amounts $B_1$, $B_2$, $\dots$, $B_M$, respectively. This system model is shown in Fig.~\ref{fig:bit_arri}. Our objective is again to minimize the transmission completion time, $T$, which again is the time by which the last bit is delivered to the destination.

Let us denote the sequence of transmission powers by $p_1$, $p_2$, $\ldots$, $p_N$, and the corresponding transmission durations by $l_1$, $l_2$, $\ldots$, $l_N$. Then, the optimization problem becomes:
\begin{eqnarray}
\min_{\pv,\lv} & &T \nonumber \\
\mbox{s.t.} & & E(t)\leq\sum_{i:s_i< t} E_i, \quad  0\leq t\leq T \nonumber \\
& &B(t)\leq\sum_{i:t_i< t} B_i, \quad 0\leq t\leq T \nonumber \\
& & B(T)=\sum_{i=0}^M B_i
\label{opt_prob2}
\end{eqnarray}
where $E(t)$, $B(t)$ are defined in (\ref{eqn:E}) and (\ref{eqn:D}). We again determine the properties of the optimal transmission policy in the following three lemmas.

\begin{Lemma}\label{lemma:incrs2}
Under the optimal solution, the transmission rates increase in time, i.e., $r_1\leq r_2 \leq \cdots \leq r_N$.
\end{Lemma}
\begin{Proof}
First, note that since the relationship between power and rate, $r=g(p)$, is monotone, stating that the rates increase monotonically is equivalent to stating that the powers increase monotonically.
 We follow steps  similar to those in the proof of Lemma~\ref{lemma:incrs} to prove this lemma. Assume that the rates do not increase monotonically, i.e., that we can find two rates such that $r_{i}>r_{i+1}$, with duration $l_i$, $l_{i+1}$, respectively. If $i+1\neq N$, then, let
\begin{align}
    r_i'&=r'_{i+1}=\frac{r_il_i+r_{i+1}l_{i+1}}{l_i+l_{i+1}}\\
    p_i'&=p'_{i+1}=g^{-1}\left(\frac{r_il_i+r_{i+1}l_{i+1}}{l_i+l_{i+1}}\right)
\end{align}
Since $r_i>r_i'=r_{i+1}'>r_{i+1}$, $p_i>p_i'=p_{i+1}'>p_{i+1}$, it is easy to verify that the new policy is feasible up to the end of $l_{i+1}$, from both the data and energy arrival points of view.  On the other hand, based on the convexity of $g^{-1}$, the energy spent over the duration $l_i+l_{i+1}$ is smaller than $p_il_i+p_{i+1}l_{i+1}$. If we allocate the saved energy over to the last transmission duration, without conflicting any energy or data constraints, the transmission will be completed in a shorter duration. If $i+1=N$, then, we let
\begin{align}
p_i'&=p'_{i+1}=\frac{p_il_i+p_{i+1}l_{i+1}}{l_i+l_{i+1}}\\
r_i'&=r'_{i+1}=g\left(\frac{p_il_i+p_{i+1}l_{i+1}}{l_i+l_{i+1}}\right)
\end{align}
Then, from (\ref{eqn:incr}), under the new policy, the last bit will depart before the end of $l_{i+1}$. The energy and data arrival constraints are satisfied over the whole transmission duration. Consequently, the original policy could not be optimal. Therefore, the optimal policy must have monotonically increasing rates (and powers).
\end{Proof}

\begin{Lemma}\label{lemma:const2}
The transmission power/rate remains constant between two event epoches, i.e., the rate only potentially changes when new energy is harvested or a new packet arrives.
\end{Lemma}
\begin{Proof}
  This lemma can be proved through a procedure similar to that in Lemma~\ref{lemma:const}. If power/rate is not constant between two event epoches, then, by equalizing the rate over the duration while keeping the total departures fixed, we can save energy. Allocating this saved energy to the last transmission duration, we can shorten the whole transmission duration. Thus, if power/rate is not constant between two event epoches, the policy cannot be optimal.
\end{Proof}

\begin{Lemma}\label{lemma:energy2}
If the transmission rate changes at an energy harvesting epoch, then the energy consumed up to that epoch equals the energy harvested up to that epoch; if the transmission rate changes at a packet arrival epoch, then, the number of packets departed up to that epoch equals the number of packets arrived up to that epoch; if the event epoch has both energy and data arrivals at the same time, then, one of the causality constraints must be met with equality.
\end{Lemma}
\begin{Proof}
This lemma can be proved through contradiction using techniques similar to those used in the proof of Lemma~\ref{lemma:energy}. When the transmission rate changes at an energy harvesting epoch, if the energy consumed up to that time is not equal to the total amount harvested, then, without conflicting the energy causality constraint, we can always increase the rate before that epoch a little and decrease the rate after that epoch a little while keeping the total departures fixed. This policy would save some energy which can be used to shorten the transmission durations afterwards. Thus, the energy constraint at that epoch must be satisfied as an equality. The remaining situations can be proved similarly.
\end{Proof}

Based on Lemmas~\ref{lemma:incrs2},~\ref{lemma:const2} and ~\ref{lemma:energy2}, we can identify the structure of the unique optimal transmission policy as stated in the following theorem. In order to simplify the notation, we define $u_i$ to be the time epoch when the $i$th arrival (energy or data) happens, i.e.,
\begin{align}
  u_1&=\min\{s_1, t_1\}\\
  u_2&=\min\{s_i,t_j:s_i>u_1, t_j>u_1\}
\end{align}
and so on, until the last arrival epoch.
\begin{Theorem}\label{thm:struct2}
For a given energy harvesting and packet arrival profile, the transmission policy with a transmission rate vector $\rv=[r_1, r_2,\ldots,r_N]$ and the corresponding duration vector $\lv=[l_1, l_2, \ldots, l_N]$ is optimal, if and only if it has the following structure:
\begin{align}
\sum_{i=1}^N r_il_i&=\sum_{i=0}^MB_i\\
  r_1&=\min_{i:u_i\leq T}\left\{g\left(\frac{\sum_{j:s_j< u_i}E_j}{u_i}\right), \frac{\sum_{j:t_j<u_i}B_j}{u_i}\right\}\label{eqn:r1}
\end{align}
Let $i_1$ be the index of $u$ associated with $r_1$. Then, with updated amount of bits and energy remaining in the system at  $t=u_{i_1}$, $r_2$ is the smallest feasible rate starting from $u_{i_1}$, and so on.
\end{Theorem}
The proof of this theorem is given in Appendix~\ref{apdix:thm3}.

For a given optimal transmission duration, $T$, the optimal policy which has the structure in Theorem~\ref{thm:struct2} is unique. However, since we do not know the exact transmission duration up front, we obtain a lower bound on $T$ first, as in the previous section. In this case also, we develop a similar procedure to find the optimal transmission policy. The basic idea is to keep the transmit power/rate as constant as possible throughout the entire transmission duration. Because of the additional casuality constraints due to data arrivals, we need to consider both the average data arrival rate as well as the average power the system can support for feasibility.

If $s_K\leq t_M$, i.e., bits have arrived after the last energy harvest, then, all of the harvested energy will be used. First, we assume that we can use these energies to maintain a constant rate, and the transmission duration will be the solution of
\begin{align}
   g\left(\frac{\sum_{j=0}^{K}E_j}{T}\right)T&=\sum_{j=0}^{M}B_j
\end{align}
Then, we check whether this constant power/rate is feasible. We check the availability of the energy, as well as the available number of bits. Let
\begin{align}
i_{1e}&=\arg\min_{u_i< T} \left\{\frac{\sum_{j=0}^{i-1}E_j}{u_i}\right\},\quad  p_{1}=\frac{\sum_{j=0}^{i_{1e}-1}E_j}{u_i}\\
i_{1b}&=\arg \min_{u_i< T} \left\{\frac{\sum_{j=0}^{i-1}B_j}{u_i}\right\},\quad
r_1=\frac{\sum_{j=0}^{i_{1b}-1}B_j}{u_i}
\end{align}
We compare $\min(p_1, g^{-1}(r_1))$ with $\frac{\sum_{j=0}^{K}E_j}{T}$. If the former is greater than the latter, then the constant transmit power $\frac{\sum_{j=0}^{K}E_j}{T}$ is feasible. Thus, we achieve the minimum possible transmission completion time $T$. Otherwise, constant-power transmission is not feasible. We choose the transmit power to be the smaller of $p_1$ and $g^{-1}(r_1)$, and the duration to be the one associated with the smaller transmit power. We repeat this procedure until all of the bits are transmitted.

If $s_K>t_M$, then, as in the first scenario where packets have arrived and are ready before the transmission starts, some of the harvested energy may not be utilized to transmit the bits. In this case also, we need to get a lower bound for the final transmission completion time. Let $u_n$ be the energy harvesting epoch right after $t_M$. Then, starting from $u_n$, we compute the energy required to transmit $\sum_{j=0}^{M}B_j$ bits at a constant rate by $u_i$, $u_n\leq u_i\leq u_{K+M}$, and compare them with the total energy harvested up to that epoch, i.e., $\sum_{j:s_j<u_i}E_j$. We identify the smallest $i$ such that the required energy is smaller than the total harvested energy, and denote it by $\tilde{i}_1$. If no such $\tilde{i}_1$ exists, we let $\tilde{i}_1=M+K+1$.

Now, we assume that we can use $\sum_{j:s_j<u_{\tilde{i}_1}}E_j$ to transmit $\sum_{j=0}^{M}B_j$ bits at a constant rate. We allocate the energy evenly to these bits, and the overall transmission time $T_1$ is the solution of
\begin{align}
 g\left(\frac{\sum_{j:s_j<u_{\tilde{i}_1}}E_j}{T_1}\right)T_1&=\sum_{j=0}^{M}B_j
\end{align}
and the corresponding constant transmit power is
\begin{align}
 p_1&=\frac{\sum_{j:s_j<u_{\tilde{i}_1}}E_j}{T_1}
\end{align}
Next, we compare $p_1$ with $\frac{\sum_{j:s_j<u_i}E_j}{u_i}$ and $g^{-1}\left(\frac{\sum_{j:t_j<u_i}B_j}{u_i}\right)$ for every $i< \tilde{i}_1$. If $p_1$ is smaller than all of these terms, then, maintaining $p_1$ is feasible from both energy and data arrival points of view. The optimal policy is to keep a constant transmission rate at $g(p_1)$ with duration $T_1$, which yields the smallest possible transmission completion time,  $i_1=\tilde{i}_1$.
Otherwise, maintaining $p_1$ is not feasible under the given energy and data arrival realizations. This infeasibility is due to the causality constraints on either the energy or the data arrival, or both. Next, we identify the tightest constraint, and update the transmit power to be the power associated with that constraint. We repeat this procedure until all of the bits are delivered.

\begin{Theorem}\label{thm4}
The transmission policy obtained through the algorithm described above is optimal.
\end{Theorem}
The proof of this theorem is given in Appendix~\ref{apdix:thm4}.

\section{Simulation Results}
\begin{figure}[h]
\begin{center}
\scalebox{0.7} {\epsffile{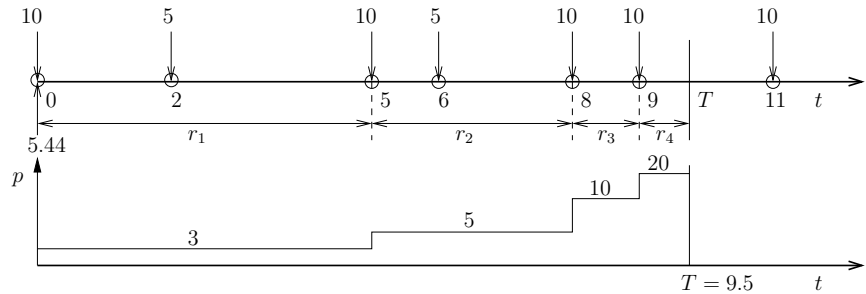}}
\end{center}
\caption{Optimal transmit powers $\pv=[3,5,10,20]$mW, with durations $\lv=[5,3,1,0.5]$s.} \label{fig:example}
\end{figure}

We consider a band-limited additive white Gaussian noise channel, with bandwidth $W=1$MHz and the noise power spectral density $N_0=10^{-19}$W/Hz. We assume that the distance between the transmitter and the receiver is 1km, and the path loss is about $110$dB. Then, we have $g(p)=W\log_2\left(1+\frac{ph}{N_0W}\right)=\log_2\left(1+\frac{p}{10^{-2}}\right)$Mbps. It is easy to verify that this function has the properties assumed at the beginning of Section~\ref{sec:no_arri}. For the energy harvesting process, we assume that at times $\tv=[0,2,5,6,8,9,11]$s, we have energy harvested with amounts $\Ev=[10,5,10,5,10,10,10]$mJ, as shown in Fig.~\ref{fig:example}. We assume that at $t=0$, we have $5.44$Mbits to transmit. We choose the numbers in such a way that the solution is expressable in simple numbers, and can be potted conveniently. Then, using our algorithm, we obtain the optimal transmission policy, which is shown in Fig.~\ref{fig:example}. We note that the powers change only potentially at instances when energy arrives (Lemma~\ref{lemma:const}); when the power changes, energy consumed up to that point equals energy harvested (Lemma~\ref{lemma:energy}); and power sequence is monotonically increasing (Lemma~\ref{lemma:incrs}). We also note that, for this case, the active transmission is completed by time $T=9.5$s, and the last energy harvest at time $t=11$s is not used.

\begin{figure}[h]
\begin{center}
\scalebox{0.7} {\epsffile{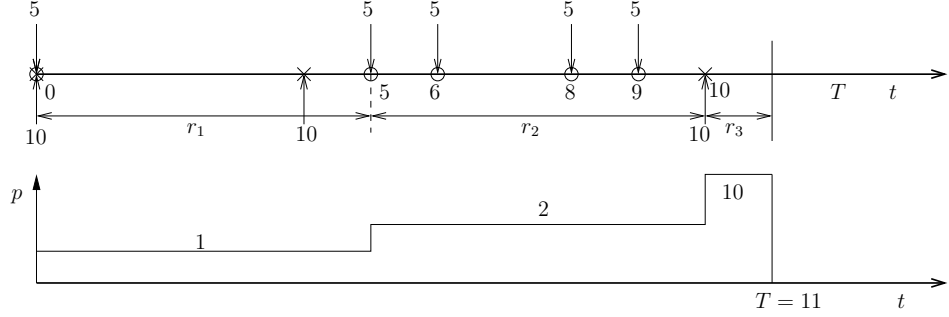}}
\end{center}
\caption{Optimal transmit powers $\pv=[1,2,10]$mW, with durations $\lv=[5,5,1]\times 10^{-2}$s. } \label{fig:example2}
\end{figure}

Next, we consider the scenario where data packets arrive during the transmissions. We consider a smaller time scale, where each unit consists of $10$ms. We assume that at times $\tv=[0, 5, 6, 8, 9]$, energies arrive with amounts $\Ev=[5,5,5,5,5]\times 10^{-2}$mJ, while at times $\tv=[0,4,10]$, packets arrive with equal size $10$kbits,  as shown in Fig.~\ref{fig:example2}. We observe that the transmitter changes its transmission power during the transmissions. The first change happens at $t=5$ when energy arrives, and the energy constraint at that instant is satisfied with equality, while the second change happens at $t=10$ when new bits arrive, and the traffic constraint at that time is satisfied with equality.

\section{Conclusions}
In this paper, we investigated the transmission completion time minimization problem in an energy harvesting communication system. We considered two different scenarios, where in the first scenario, we assume that packets have already arrived and are ready to be transmitted at the transmitter before the transmission starts, and in the second scenario, we assume that packets may arrive during the transmissions. We first analyzed the structural properties of the optimal transmission policy, and then developed an algorithm to obtain a globally optimal off-line scheduling policy, in each scenario.

\appendix
\subsection{The Proof of Theorem~\ref{thm:structure}}\label{apdix:thm1}
We will prove the necessariness and the sufficiency of the stated structure separately. First, we prove that the optimal policy must have the structure given above. We prove this through contradiction. Assume that the optimal policy, which satisfies Lemmas~\ref{lemma:incrs}, \ref{lemma:const} and \ref{lemma:energy}, does not have the structure given above. Specifically, assume that the optimal policy over the duration $[0,s_{i_{n-1}})$ is the same as the policy described in Theorem 1, however, the transmit power right after $s_{i_{n-1}}$, which is $p_n$, is not the smallest average power possible starting from $s_{i_{n-1}}$, i.e., we can find another $s_{i'}\leq s_{i_N}$, such that
\begin{align}\label{eqn:pn}
    p_n&>\frac{\sum^{{i'}-1}_{j=i_{n-1}}E_j}{s_{i'}-s_{i_{n-1}}}\triangleq p'
\end{align}
Based on Lemma~\ref{lemma:energy}, the energy consumed up to $s_{i_{n-1}}$ is equal to $\sum^{i_{n-1}-1}_{j=0}E_j$, i.e., there is no energy remaining at $t=s_{i_{n-1}}^-$.

\begin{figure}[h]
\subfigure[$s_{i'}<s_{i_n}$]{
\scalebox{0.7}{\epsffile{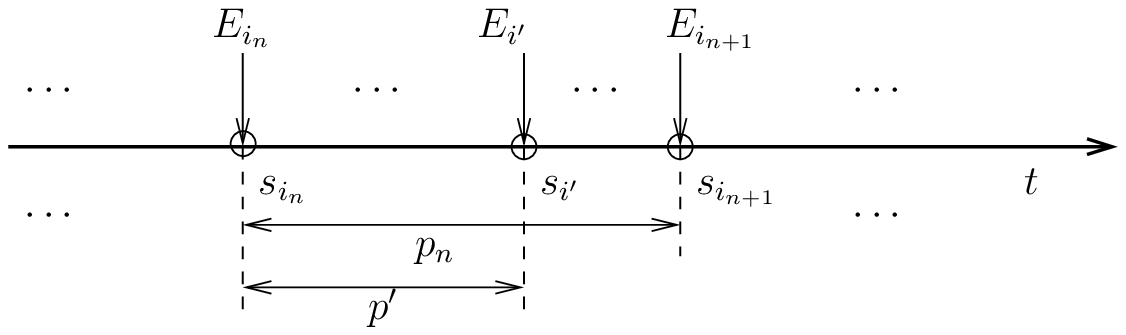}}
\label{fig:thm1_1}
}
\subfigure[ $s_{i'}>s_{i_n}$]{
\scalebox{0.7}{\epsffile{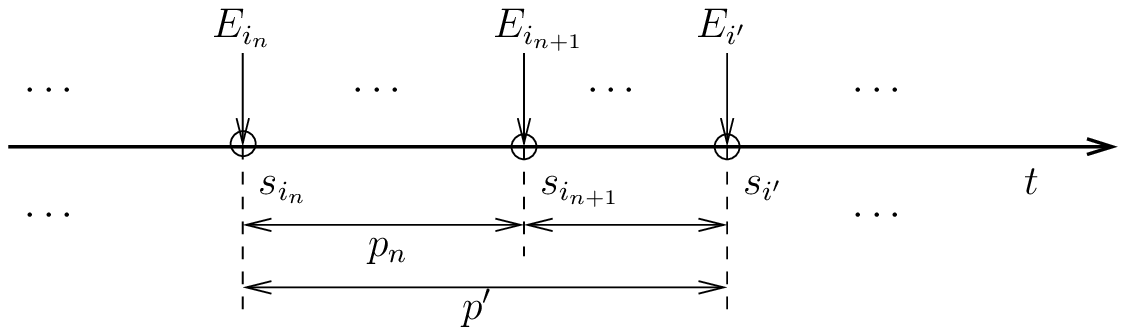}}
\label{fig:thm1_2}
}\caption{Two different cases in the proof of Theorem 1.}
\end{figure}

We consider two possible cases here. The first case is that $s_{i'}<s_{i_n}$, as shown in Fig.~\ref{fig:thm1_1}. Under the optimal policy, the energy required to maintain a transmit power $p_n$ over the duration $[s_{i_{n-1}}, s_{i'})$ is $p_n(s_{i'}-s_{i_{n-1}})$.  Based on (\ref{eqn:pn}), this is greater than the total amount of energy harvested during $[s_{i_{n-1}}, s_{i'})$, which is $\sum^{{i'}-1}_{j=i_{n-1}}E_j$. Therefore, this energy allocation under this policy is infeasible.

On the other hand, if $s_{i'}>s_{i_n}$, as shown in Fig.~\ref{fig:thm1_2}, then the total amount of energy harvested  over $[s_{i_n}, s_{i'})$ is $\sum^{{i'}-1}_{j=i_{n}}E_j$. From (\ref{eqn:pn}), we know
\begin{align}
p_n&=\frac{\sum^{{i_n}-1}_{j=i_{n-1}}E_j}{s_{i_n}-s_{i_{n-1}}}>\frac{\sum^{{i'}-1}_{j=i_{n-1}}E_j}{s_{i'}-s_{i_{n-1}}}>\frac{\sum^{{i'}-1}_{j=i_{n}}E_j}{s_{i'}-s_{i_{n}}}
\end{align}
Thus, under any feasible policy, there must exist a duration $l\subseteq [s_{i_{n}},s_{i'})$, such that the transmit power over this duration is less than $p_n$. This contradicts with Lemma~\ref{lemma:incrs}. Therefore, this policy cannot be optimal.

Next, we prove that if a policy with power vector $\pv$ and duration vector $\lv$ has the structure given above, then, it must be optimal. We prove this through contradiction. We assume that there exists another policy with power vector $\pv'$ and duration vector ${\lv}'$, and the transmission completion time $T'$ under this policy is smaller.

We assume both of the policies are the same over the duration $[0,s_{i_{n-1}})$, however, the transmit policies right after $s_{i_{n-1}}$, which are $p_n$ and $p_n'$, with durations $l_n$ and $l_n'$, respectively, are different. Based on the assumption, we must have $p_n<p_n'$.

If $l_n<l_n'$, from Lemma~\ref{lemma:energy}, we know that the total energy available over $[s_{i_{n-1}},s_{i_{n}})$ is equal to $p_nl_n$. Since $p_n<p_n'$, $p_n'$ is infeasible over $[s_{i_{n-1}},s_{i_{n}})$. Thus, policy $\pv'$ cannot be optimal.
Then, we consider the case when $l_n>l_n'$. If $T'\geq s_{i_{n}}$, then, the total energy spent over $[s_{i_{n-1}},s_{i_{n}})$ under $\pv'$ is greater than $p_nl_n$, since $p_n'>p_n$, and $p_{n+1}'>p_{n}'$ based on Lemma~\ref{lemma:incrs}. If $T'<s_{i_{n}}$, since the power-rate function $g$ is concave, the total number of bits departed over  $[s_{i_{n-1}},s_{i_{n}})$ under $\pv$ is greater than that under $\pv'$. Thus, policy $\pv'$ cannot depart $B_0$ bits over $T'$, and it cannot be optimal.

In summary, a policy is optimal if and only if it has the structure given above, completing the proof.

\subsection{The Proof of Theorem~\ref{thm2}}\label{apdix:thm2}
Let $T$ be the final transmission duration given by the allocation procedure. Then, we have $B(T)=B_0$. In order to prove that the allocation is optimal, we need to show that the final transmission policy has the structure given in Theorem~\ref{thm:structure}. We first prove that $p_1$ satisfies (\ref{eqn:p_min}). Then, we can similarly prove that $p_2$, $p_3$, $\ldots$ satisfy (\ref{eqn:p_min}).

We know that if $T=T_1$, then it is the minimum possible transmission completion time. We know that this transmit policy will satisfy the structural properties in Theorem~\ref{thm:structure}. Otherwise, the final optimal transmission time $T$ is greater than $T_1$, and more harvested energy may need to be utilized to transmit the remaining bits. From the allocation procedure, we know that
\begin{align}
  p_1\leq \frac{\sum^{i-1}_{j=0}E_j}{s_i},  \quad\forall i< \tilde{i}_1\label{eqn:p1}
\end{align}
In order to prove that $p_1$ satisfies (\ref{eqn:p_min}), we need to show that
\begin{align}
  p_1&\leq  \frac{\sum_{j=0}^{i-1}E_j}{s_i}, \quad \forall i: s_{\tilde{i}_1}\leq s_i\leq T
\label{eqn:p11}
\end{align}

\begin{figure}[h]
\subfigure[$T\leq s_{\tilde{i}_1}$]{
\scalebox{0.7}{\epsffile{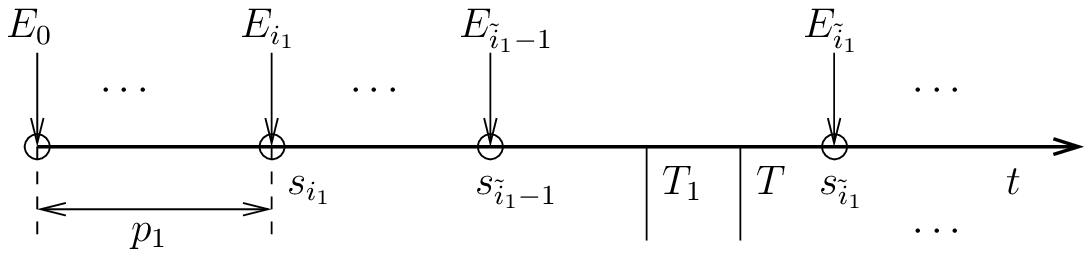}}
\label{fig:thm2_1}
}
\subfigure[ $T> s_{\tilde{i}_1}$]{
\scalebox{0.7}{\epsffile{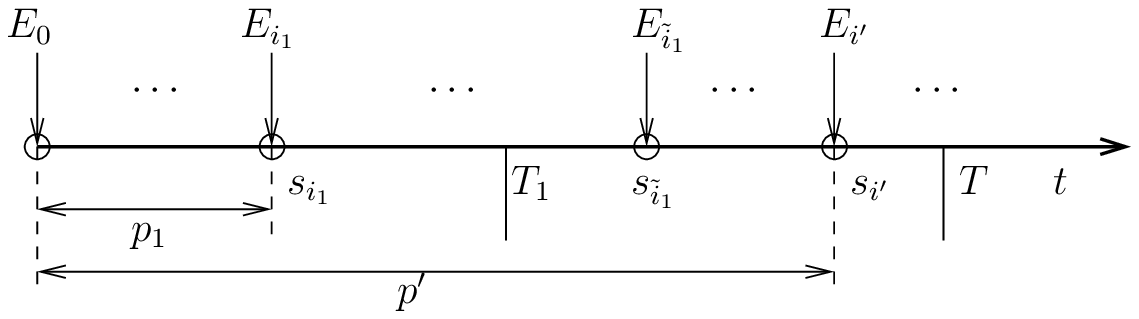}}
\label{fig:thm2_2}
}\caption{Two different cases in the proof of Theorem 2.}
\end{figure}

If we keep transmitting with power $p_1$, then at $T'_1=\frac{\sum_{j=0}^{\tilde{i}_1-1}E_j}{p_1}$, the total number of bits departed will be
\begin{align}
  g(p_1)T'_1\geq g\left(\frac{\sum_{j=0}^{\tilde{i}_1-1}E_j}{T_1}\right)T_1=B_0
\end{align}
where the inequality follows from the assumption that $g(p)/p$ decreases in $p$. Then, (\ref{eqn:p1}) guarantees that this is a feasible policy. Thus, under the optimal policy, the transmission duration $T$ will be upper bounded by $T'_1$, i.e.,
\begin{align}\label{eqn:T}
  T\leq \frac{\sum_{j=0}^{\tilde{i}_1-1}E_j}{p_1}
\end{align}
which implies
\begin{align}\label{eqn:pT}
  p_1\leq \frac{\sum_{j=0}^{\tilde{i}_1-1}E_j}{T}
\end{align}
If $T\leq s_{\tilde{i}_1}$, as shown in Fig.~\ref{fig:thm2_1}, no future harvested energy is utilized for the transmissions. Then, (\ref{eqn:pT}) guarantees that (\ref{eqn:p11}) is satisfied.

If $T>s_{\tilde{i}_1}$, as shown in Fig.~\ref{fig:thm2_2}, additional energy harvested after $s_{\tilde{i}_1}$ should be utilized to transmit the data. We next prove that (\ref{eqn:p11}) still holds through contradiction. Assume that there exists $i'$ with $s_{\tilde{i}_1}\leq s_{i'}\leq T$, such that (\ref{eqn:p11}) is not satisfied, i.e.,
\begin{align}
   p_1>\frac{\sum_{j=0}^{i'-1} E_j}{s_{i'}}\triangleq p'
\end{align}
Then,
\begin{align}
  \frac{\sum_{j=0}^{i'-1} E_j}{p_1}&< s_{i'}
\end{align}
Combining this with (\ref{eqn:T}), we have $T<s_{i'}$, which contradicts with the assumption that $s_{i'}\leq T$. Thus, (\ref{eqn:p11}) holds, $p_1$ satisfies the requirement of the optimal structure in (\ref{eqn:p1}).

We can then prove using similar arguments that $p_2$, $p_3$, $\ldots$ also satisfy the properties of the optimal solution. Based on Lemma~\ref{thm:structure}, this procedure gives us the unique optimal policy.

\subsection{The Proof of Theorem~\ref{thm:struct2}}\label{apdix:thm3}
  First, we prove that for the optimal transmission policy, $r_1$ must satisfy (\ref{eqn:r1}). We prove this through contradiction. If $r_1$ does not satisfy (\ref{eqn:r1}), then, we can always find another $u_{i'}$, such that
  \begin{align}
    r_1&>\min \left\{g\left(\frac{\sum_{j:s_j< u_{i'}}E_j}{u_{i'}}\right), \frac{\sum_{j:t_j<u_{i'}}B_j}{u_{i'}}\right\}
  \end{align}
   First, we assume that $g\left(\frac{\sum_{j:s_j< u_{i'}}E_j}{u_{i'}}\right)<\frac{\sum_{j:t_j<u_{i'}}B_j}{u_{i'}}$. Then, if $u_{i'}<u_{i_1}$, clearly $r_1$ is not feasible over the duration $[0,u_{i'})$, because of the energy constraint. If $u_{i'}>u_{i_1}$, then, the transmitter cannot maintain a transmission rate that is always greater than $r_1$ over $[u_{i},u_{i'})$, from the energy point of view. This contradicts with Lemma~\ref{lemma:incrs2}. Similarly, if $g\left(\frac{\sum_{j:s_j< u_{i'}}E_j}{u_{i'}}\right)>\frac{\sum_{j:t_j<u_{i'}}B_j}{u_{i'}}$, the ``bottleneck'' is the data constraint. We can prove that $r_1$ is not feasible. Thus, $r_1$ must be the smallest feasible rate starting from $t=0$, as in (\ref{eqn:r1}). We can also prove that $r_2$, $r_3$, $\ldots$ must have the same structure, in the same way. Next, we can prove that any policy has the structure described above is optimal. We can prove this through contradiction. Assume that there exists another policy with a shorter transmission completion time. Based on Lemmas~\ref{lemma:incrs2} and ~\ref{lemma:energy2}, we can prove that this policy could not be feasible.

 \subsection{The Proof of Theorem~\ref{thm4}}\label{apdix:thm4}
  First we prove that $r_1$ obtained through this procedure satisfies (\ref{eqn:r1}). If $T=T_1$, i.e., the constant rate is achievable throughout the transmission, then it is the shortest transmission duration we can get, thus, it is optimal. If $T\neq T_1$, from the procedure, we have
  \begin{align}
    r_1&\leq \min_{1\leq i\leq \tilde{i}_1}\left\{g\left(\frac{\sum_{j:s_j< u_i}E_j}{u_i}\right), \frac{\sum_{j:t_j<u_i}B_j}{u_i}\right\}
  \end{align}
  We need to prove that
  \begin{align}\label{eqn:r11}
     r_1&\leq \min \left\{g\left(\frac{\sum_{j:s_j< u_i}E_j}{u_i}\right), \frac{\sum_{j:t_j<u_i}B_j}{u_i}\right\}\quad \mbox{for } u_{\tilde{i}_1}<u_i\leq T.
  \end{align}
  Considering the policy with a constant power $p_1=g^{-1}(r_1)$, then, at $T'_1=\frac{\sum_{j=0}^{\tilde{i}_1-1}E_j}{p_1}$, the total number of bits departed will be
\begin{align}
  g(p_1)T'_1\geq g\left(\frac{\sum_{j=0}^{\tilde{i}_1-1}E_j}{T_1}\right)T_1=\sum_{j=0}^{M}B_j
\end{align}
while at $T''_1=\frac{\sum_{j=0}^{\tilde{i}_1-1}B_j}{g(r_1)}$, the total energy required will be
\begin{align}
  p_1T''_1\leq \frac{\sum_{j=0}^{\tilde{i}_1-1}E_j}{T_1}\frac{\sum_{j=0}^{\tilde{i}_1-1}B_j}{g\left(\frac{\sum_{j=0}^{\tilde{i}_1-1}E_j}{T_1}\right)}=\sum_{j=0}^{\tilde{i}_1-1}E_j
\end{align}
where the inequality follows from the assumption that $g(p)/p$ decreases in $p$. Therefore, maintaining a transmission rate $r_1$ until the last bit departs the system is feasible from both the energy and data arrival points of view. Thus, under the optimal policy, the transmission duration $T$ will be upper bounded by $T'_1$ and $T''_1$, i.e.,
\begin{align}\label{eqn:T2}
  T\leq \frac{\sum_{j=0}^{\tilde{i}_1-1}E_j}{p_1}, \quad T\leq \frac{\sum_{j=0}^{\tilde{i}_1-1}B_j}{r_1}
\end{align}
which implies
\begin{align}\label{eqn:pT2}
  p_1\leq \frac{\sum_{j=0}^{\tilde{i}_1-1}E_j}{T}, \quad r_1\leq  \frac{\sum_{j=0}^{\tilde{i}_1-1}B_j}{T}
\end{align}
If no future harvested energy is utilized for the transmissions, (\ref{eqn:pT2}) guarantees that (\ref{eqn:r1}) is satisfied.

If $T>u_{\tilde{i}_1}$, additional energy harvested after $u_{\tilde{i}_1}$ should be utilized to transmit the data. We next prove that (\ref{eqn:r11}) still holds through contradiction. Assume that there exists $i'$ with $u_{\tilde{i}_1}\leq u_{i'}\leq T$, such that (\ref{eqn:r11}) is not satisfied, i.e.,
\begin{align}
   p_1>\frac{\sum_{j=0}^{i'-1} E_j}{u_{i'}}\quad \mbox{or} \quad r_1>\frac{\sum_{j=0}^{i'-1} B_j}{u_{i'}}
\end{align}
Then, we have
\begin{align}
  \frac{\sum_{j=0}^{i'-1} E_j}{p_1}< u_{i'} \quad \mbox{or} \quad  \frac{\sum_{j=0}^{i'-1} B_j}{r_1}< u_{i'}
\end{align}
Combining this with (\ref{eqn:T2}), we have $T<u_{i'}$, which contradicts with the assumption that $u_{i'}\leq T$. Thus, (\ref{eqn:r11}) holds, $r_1$ satisfies the requirement of the optimal structure in (\ref{eqn:r1}). We can then prove using a similar argument that $r_2$, $r_3$, $\ldots$ also satisfy the structure of the optimal solution. Based on Theorem~\ref{thm:struct2}, this procedure gives us the unique optimal transmission policy.

\end{document}